\documentclass[aps,preprintnumbers,eqsecnum,amsmath,amssymb,showpacs,nofootinbib]{revtex4}
\usepackage{graphicx}
\usepackage{dcolumn}
\usepackage{bm}
\usepackage{epsfig}

\begin{document} 

\preprint{HEPHY-PUB 973/16} 
\preprint{RM3-TH/16-9}
\vspace*{0.5cm}

\title{\boldmath Isospin breaking in the decay constants of heavy mesons from QCD sum rules}
\author{Wolfgang Lucha$^a$, Dmitri Melikhov$^{a,b,c}$, Silvano Simula$^d$}
\affiliation{$^a$Institute for High Energy Physics, Austrian Academy of Sciences, Nikolsdorfergasse 18, 
A-1050 Vienna, Austria\\
$^b$D.~V.~Skobeltsyn Institute of Nuclear Physics,
M.~V.~Lomonosov Moscow State University, 119991, Moscow, Russia\\
$^c$Faculty of Physics, University of Vienna, Boltzmanngasse 5, A-1090 Vienna, Austria\\ 
$^d$INFN, Sezione di Roma III, Via della Vasca Navale 84, I-00146 Roma, Italy}

\begin{abstract}
We present a study of the strong isospin-breaking (IB) effect, due in QCD to the difference between 
$u$- and $d$-quark masses, in the leptonic decay constants of charmed and beauty pseudoscalar and vector 
mesons using the method of QCD sum rules. 
We apply the sum-rule analysis to the decay constants of mesons containing one heavy quark and one 
light quark with the light mass in the range from the average $u/d$ quark mass to the strange-quark mass.
We then analyse the dependence of the decay constants on the light-quark mass and extract with good accuracy 
the IB ratios of decay constants at leading order in the mass difference $(m_d - m_u)$, obtaining: 
$(f_{D^+} - f_{D^0}) / f_{D} = 0.0047 (6)$,
$(f_{D^{*+}} - f_{D^{*0}}) / f_{D^*} = 0.0068 (9)$, 
$(f_{B^0} - f_{B^+}) / f_{B} = 0.0047 (6)$, and
$(f_{B^{*0}} - f_{B^{*+}}) / f_{B^*} = 0.0045 (5)$, which yield:
$f_{D^+} - f_{D^0} = 0.97 \pm 0.13$ MeV, 
$f_{D^{*+}} - f_{D^{*0}} = 1.73 \pm 0.27$ MeV,
$f_{B^0} - f_{B^+} = 0.90 \pm 0.13$ MeV, 
$f_{B^{*0}} - f_{B^{*+}} = 0.81 \pm 0.11$ MeV.
In the case of the $D$-meson our finding is consistent with recent lattice QCD results, 
whereas it is much lower in the case of the $B$-meson showing a tension of $\approx 3$ standard deviations.
\end{abstract}


\maketitle

\section{Introduction}

The QCD sum-rule approach \cite{svz,aliev,rubinstein}, based on the application of Wilson's Operator 
Product Expansion (OPE) to the properties of individual hadrons, has been extensively used for predicting 
the leptonic decay constants of heavy-light mesons. 
An important finding of these analyses was the strong sensitivity of the decay constants to the values of 
the input OPE parameters and to the prescription of fixing the effective continuum threshold \cite{lms_1}. 
The latter governs the accuracy of the quark--hadron duality approximation and, to a large extent, determines 
the extracted value of the decay constant. 
Even if the parameters of the truncated OPE are known with arbitrarily high precision, the decay constants may 
be predicted with only limited accuracy, which we refer to as their systematic uncertainty. 
In a series of papers \cite{lms_new} we have formulated a new algorithm for fixing the effective threshold within 
Borel QCD sum rules and for obtaining reliable estimates for the systematic uncertainties. 
Our procedure has opened the possibility to provide predictions for the decay constants with a controlled 
accuracy \cite{lms_fp,lms_fD+fDstar,ourmb,lms_fB_ratio_1,lms_fB_ratio} and thus to address subtle effects 
which require a solid accurate analysis. 

In particular, the application of QCD sum rules gave the first indication \cite{lms_fB_ratio_1} of the 
unexpected feature $f_{B^*}/f_B<1$, which has been confirmed subsequently by a more detailed QCD sum-rule 
analysis \cite{lms_fB_ratio} and by results from QCD simulations on the lattice \cite{hpqcd2015}. 

In this paper we discuss a different application of QCD sum rules to a subtle effect -- namely, 
the isospin breaking (IB) in the decay constants of heavy-light pseudoscalar and vector mesons 
due in QCD to the difference between $u$- and $d$-quark masses. 
Our analysis takes advantage of the fact that the OPE provides the analytic dependence of the correlation 
functions on the light-quark mass and thus allows one to study the impact of this mass on the decay constants. 

The leptonic decay constants of pseudoscalar and vector mesons are defined as  
\begin{eqnarray}
\label{f}
\langle 0|\bar q \gamma_\mu \gamma_5 Q|P_q(p)\rangle=i f_{P_q}p_\mu, \qquad \qquad 
\langle 0|\bar q \gamma_\mu Q|V_q(p)\rangle=f_{V_q}M_{V_q}\varepsilon_\mu(p) ~ , 
\end{eqnarray}
where $Q$($q$) is the heavy(light) quark.
We are interested in the IB effects on the decay constants, i.e.~in the difference between the decay 
constants of $\bar{Q}d$ and $\bar{Q}u$ mesons originating in QCD from the quark mass difference $\delta m \equiv m_d - m_u$. 

The Borel QCD sum rule for the decay constant of a heavy-light meson $H_q$ (either pseudoscalar or vector) 
consisting of a heavy quark $Q$ with mass $m_Q$ and a light quark $q$ with mass $m_q$ has the following form: 
\begin{eqnarray}
\label{srN}
f_{H_q}^2 (M_{H_q}^2)^N \exp(-M_{H_q}^2 \tau) &=&\Pi^{(N)}_{\rm dual}(\tau, s_{\rm eff}^{H_q},m_Q,m_q,m_{\rm sea}) \\
& = & \int\limits_{(m_Q+m_q)^2}^{s_{\rm eff}^{H_q}} ds ~ e^{-s \tau} s^N \rho_{\rm pert}(s,m_Q,m_q,m_{\rm sea},\alpha_s) +
          \Pi^{(N)}_{\rm power}(\tau,m_Q,m_q,\langle \bar qq \rangle,...) ~ , \nonumber 
\end{eqnarray}
where $f_{H_q}$ is the decay constant of the $\bar{Q} q$-meson, $M_{H_q}$ is its mass, $\tau $ is the Borel 
parameter and $s_{\rm eff}^{H_q}$ is the effective continuum threshold. 
In Eq.~(\ref{srN}) $N$ is an integer number related to the specific Lorentz structure of the correlation 
function chosen for the sum rule (namely, $N = 1$ for vector and axial-vector currents and $N = 2$ for pseudoscalar currents), 
while $q$ denotes the ``valence'' light quark ($u$, $d$ or $s$) entering the relevant interpolating current (\ref{f}). 
There are also ``sea'' quarks which appear in the loops, and $m_{\rm sea}$ denotes the set of the sea-quark masses $\{m_u, m_d, m_s, ...\}$.  

Both the $m_q$ and $m_{\rm sea}$ dependencies can contribute to the IB effect in Eq.~(\ref{ib}).
The dependence on the valence quark mass $m_q$ contributes at leading and higher orders in $\delta m = m_d - m_u$, 
because either one valence $u$ or one valence $d$ quark is contained in the heavy-light meson, whereas the 
dependence on $m_{\rm sea}$ contributes only at second and higher orders in $\delta m$, because both $u$ and $d$ 
quarks are equally present in the sea.
Since the relevant parameter $\delta m / \Lambda_{QCD}$ is a small quantity (of the order of $10^{-2}$), 
the IB effect arising from the difference in the $u$ and $d$ sea-quark masses can be safely neglected. 

The $m_q$-dependence, and correspondingly the IB effect, in the l.h.s.~of Eq.~(\ref{srN}) is contained 
both in the decay constant $f_{H_q}$ and in the meson mass $M_{H_q}$. 
In the r.h.s.~of Eq.~(\ref{srN}), the IB effect may come from several sources: ~ i) the $m_q$-dependence 
of $\rho_{\rm pert}(s,m_Q,m_q,m_{\rm sea},\alpha_s)$; ~ ii) the (implicit or explicit) $m_q$-dependence of the 
effective threshold $s_{\rm eff}^{H_q}$; ~ iii) the $m_q$-dependence of the power corrections; ~ iv) 
the flavour-dependence of the condensates, in particular of the quark condensate $\langle \bar qq \rangle$. 

In general, all the above effects mix together making the determination of the IB effect in $f_{H_q}$ rather challenging. 
Nevertheless, QCD sum rules provide a promising possibility to study the leading IB effect, 
since the $m_q$-dependence of the OPE is known explicitly, and thus one can perform the sum-rule analysis of $f_{H_q}$ 
for different values of $m_q$.
In this way one can address the difference $(f_{H_d} - f_{H_u})$ at first order in the $u$- and $d$-quark mass 
difference $\delta m$ from the slope of $f_{H_q}$ with respect to $m_q$, namely
\begin{eqnarray}
  \label{ib}
  f_{H_d} - f_{H_u} = \left[ \frac{\partial f_{H_q}(m_q)}{\partial m_q} \right]_{m_q = m_{ud}} \delta m + O(\delta m^2) ~ ,
\end{eqnarray}
where $m_{ud} \equiv (m_u + m_d) / 2$ is the average $u/d$ quark mass (see also Ref.~\cite{lattice_IB}).

Notice that the perturbative spectral density appearing in Eq.~(\ref{srN}) is obtained as an expansion in powers of $a=\alpha_s/\pi$: 
\begin{eqnarray}
\label{rho}
\rho_{\rm pert}(s,m_Q,m_q, m_{\rm sea}, \alpha_s) = \rho^{(0)}(s, m_Q, m_q) + a \rho^{(1)}(s, m_Q, m_q) + a^2 \rho^{(2)}(s, m_Q, m_q, m_{\rm sea}) + \dots
\end{eqnarray}
Let us emphasize that the sea-quark mass contribution starts to appear in the perturbative spectral density only at order $a^2$. 
Whereas the full $m_q$-dependence of the LO and NLO spectral densities $\rho^{(0,1)}(s,m_Q,m_q)$ is known, the NNLO spectral 
density has been calculated in Ref.~\cite{chetyrkin} only for massless light valence and sea quarks, namely $\rho^{(2)}(s,m_Q,m_q=0,m_{\rm sea}=0)$. 
Such an approximate knowledge of $\rho^{(2)}$ yields however an error of order $O(a^2 \delta m)$ in the difference (\ref{ib}), 
since the sea-quark mass effects cancel each other. 

Thus, in order to determine the leading IB effect (\ref{ib}) we propose the following strategy:
\begin{itemize}
\item consider the decay constant $f_{H_q}$, corresponding to the correlation function in which the light-quark mass 
in the LO and the NLO spectral densities is equal to $m_q$, for various values of $m_q$ chosen in the range $m_{ud} < m_q < m_s$, 
whereas in the NNLO spectral density the $u$, $d$ and $s$ quarks are considered massless;
\item parameterize the $m_q$-dependencies of the meson mass $M_{H_q}$ and of the condensate $\langle \bar qq \rangle$;
\item perform the Borel sum-rule analysis for the ratio $R_{H_q}(m_q) = f_{H_q}(m_q) / f_{H_{ud}}(m_{ud})$, where $H_{ud}$ is the 
heavy meson containing a light quark with mass equal to $m_{ud}$, using our algorithm based on the $\tau$-dependent effective 
threshold $s_{\rm eff}^{H_q}(\tau)$, which has been successfully applied to the case of the decay constants of heavy-light mesons 
\cite{lms_fp,lms_fD+fDstar,ourmb,lms_fB_ratio_1,lms_fB_ratio};
\item calculate numerically the slope of $R_{H_q}(m_q)$ at $m_q = m_{ud} \equiv (m_d + m_u) /2$ and multiply it by the value of the 
light-quark mass difference $\delta m = m_d - m_u$, taken from the updated FLAG \cite{FLAG} or PDG \cite{PDG} reviews, 
in order to get the quantity $(f_{H_d} - f_{H_u}) / f_{H_{ud}}$.
\end{itemize}
Notice again that using an approximate OPE (i.e.~massive light quarks at order $O(1)$ and $O(a)$ and massless light 
quarks at order $O(a^2)$) compared to the ``real'' OPE (massive light quarks at all orders of the perturbative expansion) 
leads to an OPE-induced error of order $O(a^2 m_s)$ for the individual decay constants, whereas the error for the 
difference of the decay constants in Eq.~(\ref{ib}) is $O(a^2 \delta m) $, which allows us to address properly the leading IB effect.

\section{Borel sum rules for the decay constants of charmed and beauty mesons}
 
We perform a sum-rule analysis of the decay constants of both pseudoscalar and vector heavy-light mesons consisting 
of a heavy quark of mass $m_Q$ (either $c$ or $b$) and a light quark of mass $m_q$ and study the dependence of the 
decay constant on $m_q$. 
For the application  of the quark-hadron duality hypothesis we make use of our algorithm based on a $\tau$-dependent 
effective threshold \cite{lms_new,lms_fp}.
As shown in \cite{lms_new}, the $\tau$-dependence of $s_{\rm eff}^{H_q}(\tau)$ can be modelled by simple polynomial 
Ans\"atze of order $n$, namely 
\begin{eqnarray}
\label{seff}
s_{\rm eff}^{H_q}(\tau) = \sum_{j=0}^n s_j^{(n)}(m_Q,m_q) \tau^j ~ ,
\end{eqnarray}
where the parameters $s_j^{(n)}(m_Q,m_q)$ of the effective threshold $s_{\rm eff}^{H_q}(\tau)$ are fixed by requiring the 
most accurate reproduction of the 
meson mass in the chosen Borel $\tau$-window. Since the meson mass depends on the masses of the appropriate 
valence quarks, $m_Q$ and $m_q$,   
the obtained effective thresholds depend on these quark masses, too. This dependence remains however implicit 
since we obtain the threshold parameters 
by a numerical procedure. The sum rules with $\tau$-dependent threshold reproduce excellently the meson mass 
in the full $\tau$-windows considered:  
for the charmed and strange charmed mesons we refer to \cite{lms_fD+fDstar} 
(e.g. Figs.~2 and 4 in the first reference of \cite{lms_fD+fDstar} for $D$ and $D_s$ mesons), 
and for beauty and strange beauty mesons this was shown in \cite{ourmb,lms_fB_ratio}
(see, e.g. Fig.~2 in \cite{ourmb} for $B$-meson and Fig.~3 in \cite{lms_fB_ratio} for $B^*$-meson). 
The same excellent reproduction of the meson mass holds also for the light quark masses in the range of $m_q$ 
from $m_{ud}$ to $m_s$. 

Once the parameters of the effective threshold are fixed, one readily obtains the decay constants from the 
sum rule (\ref{srN}). 
The estimates for the decay constants corresponding to the $n=1$ (linear), $n=2$ (quadratic) and $n=3$ (cubic) 
Ans\"atze provide a band of values: 
the central value of this band yields the sum-rule estimate for the decay constant, whereas the half-width of 
the band characterises the 
systematic uncertainty of the sum-rule estimate. 

We use here precisely the same procedures which have been successfully applied to the analysis of both 
pseudoscalar and vector heavy-light 
mesons in Refs.~\cite{lms_fp,lms_fD+fDstar,lms_fB_ratio,lms_fB_ratio_1,ourmb}.
The explicit dependence of the perturbative spectral densities on both the heavy-quark mass $m_Q$ and 
the light-quark mass $m_q$ is known: 
we make use of the axial-vector \cite{jamin} and vector \cite{khod} correlation functions in which we 
take into account the full $m_q$-dependence 
in the leading-order perturbative spectral density, as well as in the $O(m_q^0 \alpha_s)$, $O(m_q \alpha_s)$, and $O(m_q^0 \alpha_s^2)$ 
perturbative corrections.
We make use of the running-mass OPE (given always in the $\overline{\rm MS}$ scheme), 
which provides a better convergence of the perturbative expansion for the decay constants \cite{jamin} 
compared to the pole-mass OPE \cite{chetyrkin}. 
We don't repeat here the details of our analysis and just present the inputs and the final results. 
We would just like to emphasize that 
the final IB in the decay constants is the result of the interplay between the explicit 
$m_q$-dependence of the spectral densities and the implicit $m_q$-dependence of the effective thresholds.

The numerical values adopted for the relevant OPE parameters are summarized below:
\begin{eqnarray}
\label{Table:1} 
m_{ud}(2\;{\rm GeV}) \equiv \frac{m_u + m_d}{2} & = & (3.70 \pm 0.17)\;{\rm MeV} ~ \mbox{\cite{FLAG}} ~ , \nonumber \\[2mm]
m_s(2\;{\rm GeV}) & = & (93.9 \pm 1.1)\;{\rm MeV} ~ \mbox{\cite{FLAG}} ~ , \nonumber \\[2mm]
m_b(m_b) & = & (4.247 \pm 0.034)\ {\rm GeV} ~ \mbox{\cite{ourmb}} ~ , \nonumber \\[2mm]
m_c(m_c) & = & (1.275 \pm 0.025)\;{\rm GeV} ~ \mbox{\cite{PDG}} ~ , \nonumber \\[2mm]
\alpha_{\rm s}(M_Z) & = & 0.1184 \pm 0.0020 ~ \mbox{\cite{lms_fD+fDstar,lms_fB_ratio}} ~ , \nonumber \\[2mm]
\langle \bar \ell\ell \rangle(2\;{\rm GeV}) \equiv \frac{\langle \bar uu \rangle + \langle \bar dd \rangle}{2} & = &
  -[(267 \pm 17)\;{\rm MeV}]^3 ~ \mbox{\cite{lms_fD+fDstar,lms_fB_ratio,jamin}} ~ , \nonumber \\[2mm] 
\frac{\langle \bar ss \rangle(2\;{\rm GeV})}{\langle \bar \ell\ell \rangle(2\;{\rm GeV})} & = & 0.8 \pm 0.3 ~ 
\mbox{\cite{lms_fD+fDstar,lms_fB_ratio,jamin} } , \nonumber \\[2mm]
\left\langle\frac{\alpha_{\rm s}}{\pi}GG\right\rangle & = & (0.024\pm0.012)\;{\rm GeV}^4 ~ 
\mbox{\cite{lms_fD+fDstar,lms_fB_ratio,jamin}} ~ , \nonumber \\[2mm]
\frac{\left\langle \bar \ell g_s \sigma G \ell \right\rangle(2\;{\rm GeV})}{\langle \bar \ell\ell \rangle(2\;{\rm GeV})} & = & 
(0.8 \pm 0.2)\;{\rm GeV}^2 ~ \mbox{\cite{lms_fD+fDstar,lms_fB_ratio,jamin}} ~ .
\end{eqnarray}
The values of the OPE parameters are given in (\ref{Table:1}) at their standard scales. When evaluating the sum rules one, of course, 
evolves  
all the parameters to one and the same scale making use of the known evolution properties of these parameters. The relevant scale 
for charmed mesons is taken in the range $1< \mu~({\rm GeV})<3$ \cite{lms_fD+fDstar} and for beauty mesons in the range 
$3< \mu~({\rm GeV})<5$ \cite{ourmb,lms_fB_ratio_1,lms_fB_ratio}. As shown in \cite{lms_fD+fDstar} and \cite{ourmb,lms_fB_ratio_1,lms_fB_ratio}, 
the available perturbative contributions taken into account in our analysis (LO, NLO, NNLO) exhibit a good hierarchy 
in the range of scales mentioned above and in the considered windows of the Borel parameter $\tau$; 
the higher-dimension condensates provide negligible contributions to the sum rules and may be safely omitted 
\cite{jamin,khod}. Therefore the known OPE provides the possibility of a reliable analysis of the IB effects. 

In the framework of our strategy, we have to take into account the dependences of the meson mass $M_{H_q}$ and of the quark condensate 
$\langle \bar qq \rangle$ on the light-quark mass $m_q$. 
Available results from lattice QCD simulations in the charm and bottom sectors \cite{ETMC_Nf2,ETMC_Nf211,ETMC_Nf211_b} suggest that the 
light-quark mass dependence of the (pseudoscalar) meson mass $M_{H_q}$ is approximately linear.
Therefore, in what follows we assume a simple linear interpolation between the heavy-meson mass $M_{H_{ud}}$, 
corresponding to a light quark with the average $u/d$ quark mass $m_{ud}$, and the strange heavy-meson mass $M_{H_s}$, namely
\begin{eqnarray}
\label{MHq}
M_{H_q}(x_q) = M_{H_{ud}} + x_q ~ \left[ M_{H_s} - M_{H_{ud}} \right] ~ , 
\end{eqnarray}
where the variable $x_q$, defined as
\begin{eqnarray}
\label{xq}
x_q \equiv \frac{m_q - m_{ud}}{m_s - m_{ud}} ~ ,
\end{eqnarray}
ranges from $0$ at $m_q = m_{ud}$ up to $1$ at $m_q = m_s$ and is renormalization scale independent.
We adopt Eq.~(\ref{MHq}) to describe the $m_q$-dependence of both pseudoscalar and vector heavy-light mesons 
using the PDG values \cite{PDG} for the meson masses $M_{H_{ud}}$ and $M_{H_s}$.

According to SU(3) Chiral Perturbation Theory (ChPT), light-quark condensates contain chiral logs which have been 
calculated at NLO \cite{qq-condensate}.
We have checked, however, that the inclusion of such chiral effects does not have any visible impact on the 
extracted value of the decay constants in the 
considered range $0 < x_q < 1$; in this range of $x_q$ a linear approximation 
\begin{eqnarray}
\label{qq}
\langle \bar qq \rangle = \langle \bar \ell\ell \rangle + x_q ~ \left[ \langle \bar ss \rangle - \langle \bar \ell\ell \rangle \right] ~ . 
\end{eqnarray}
may be safely used.

Having fixed all the necessary inputs, the application of our procedure yields the decay constants of pseudoscalar and vector
 mesons defined in Eq.~(\ref{f}). 
The decay constants are (renormalization) scale-independent quantities. 
Nevertheless, the outcome of the sum-rule extraction provides results which depend on the specific scale $\mu$ at which the analysis is done. 
The main origin of such an unphysical $\mu$-dependence of the decay constants is the truncation of the perturbative expansion which leads 
to spectral densities depending on the scale $\mu$. 
Another source for the $\mu$-dependence is the application of the duality cut: the effective threshold is determined at each scale $\mu$, 
separately, in order to reproduce some of the heavy-meson observables and predict in this way other observables.
In particular, we fix the threshold $s_{\rm eff}^{H_q}$ in order to reproduce the meson mass $M_{H_q}$ and then use this threshold to predict 
the decay constant. 
The application of our algorithms was shown to reduce considerably the unphysical $\mu$-dependence of the decay constants obtained from 
the sum rules \cite{lms_fp,lms_fD+fDstar,ourmb,lms_fB_ratio_1,lms_fB_ratio}. 
Nevertheless, a residual $\mu$-dependence of the sum-rule estimates for decay constants is still present. 
Therefore we present the results of our analysis at fixed values of the scale $\mu$, namely $\mu = 1.7$ GeV for $D$ and $D^*$ mesons 
and $\mu = 3.75$ GeV for $B$ and $B^*$ mesons. The residual $\mu$-dependence is then added in quadrature in the final uncertainty of 
our sum-rule estimates.

\begin{figure}[!hbt]
\begin{tabular}{cc}
\includegraphics[width=8cm]{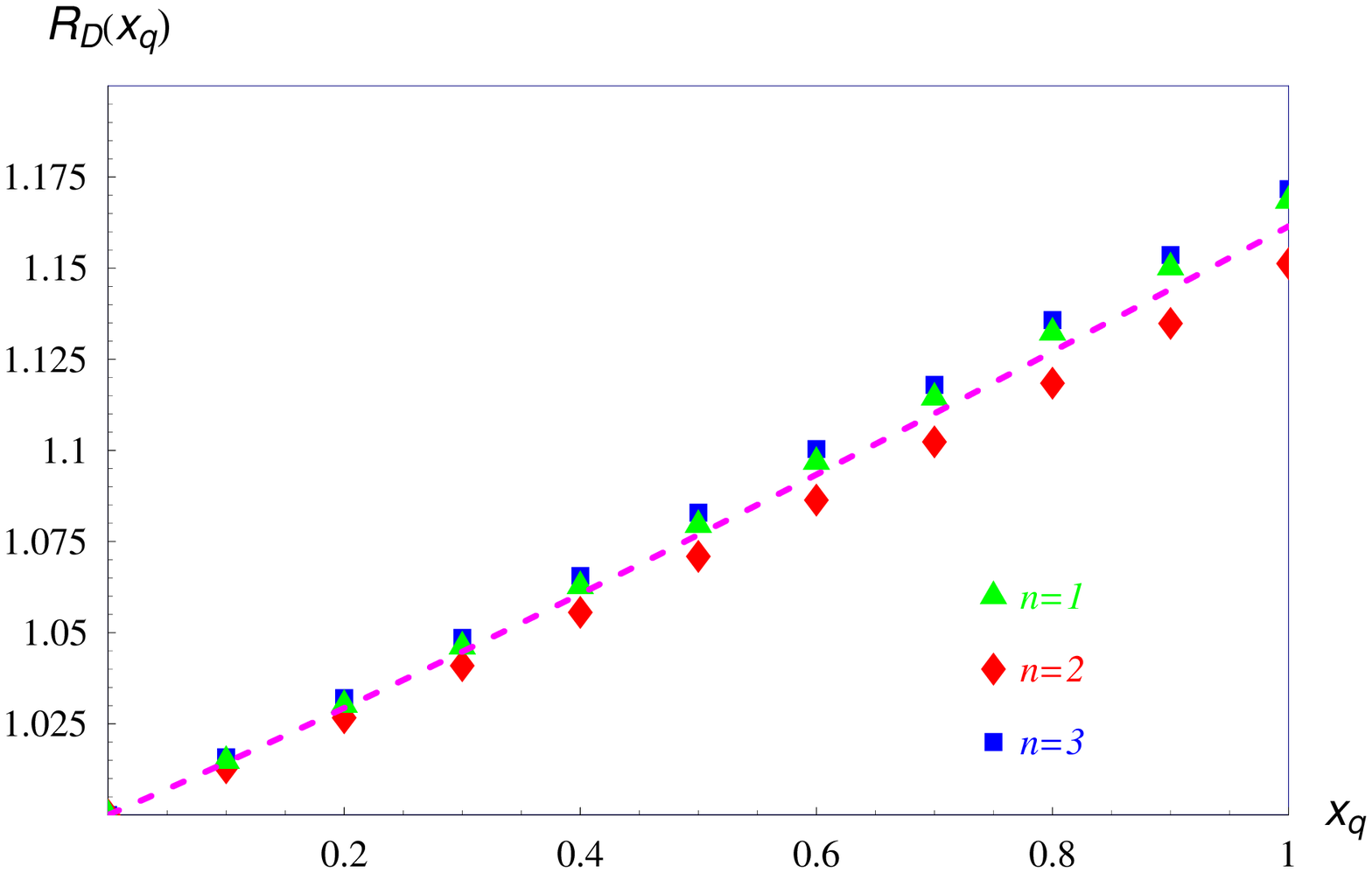} & \includegraphics[width=8cm]{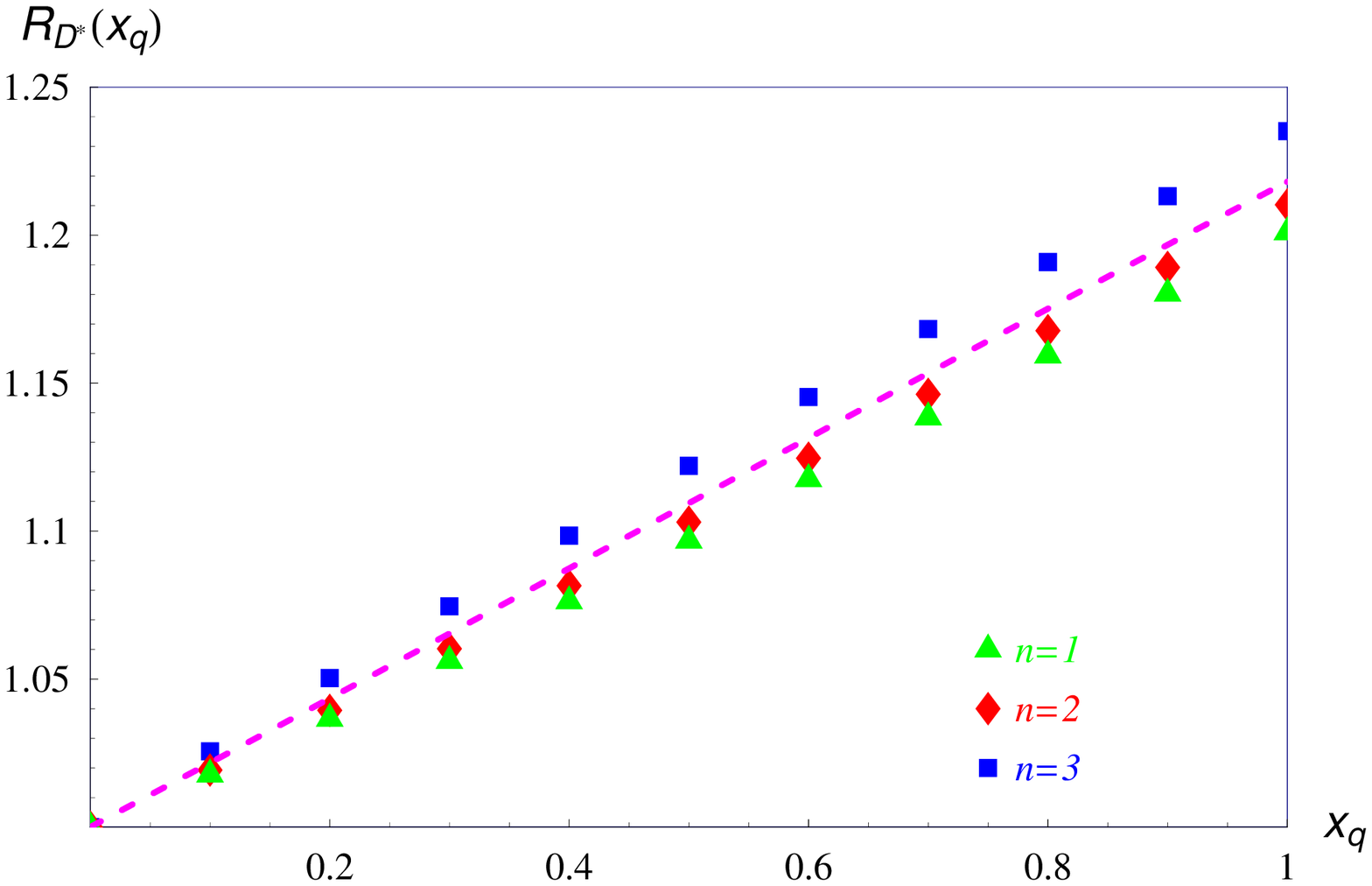} \\
\includegraphics[width=8cm]{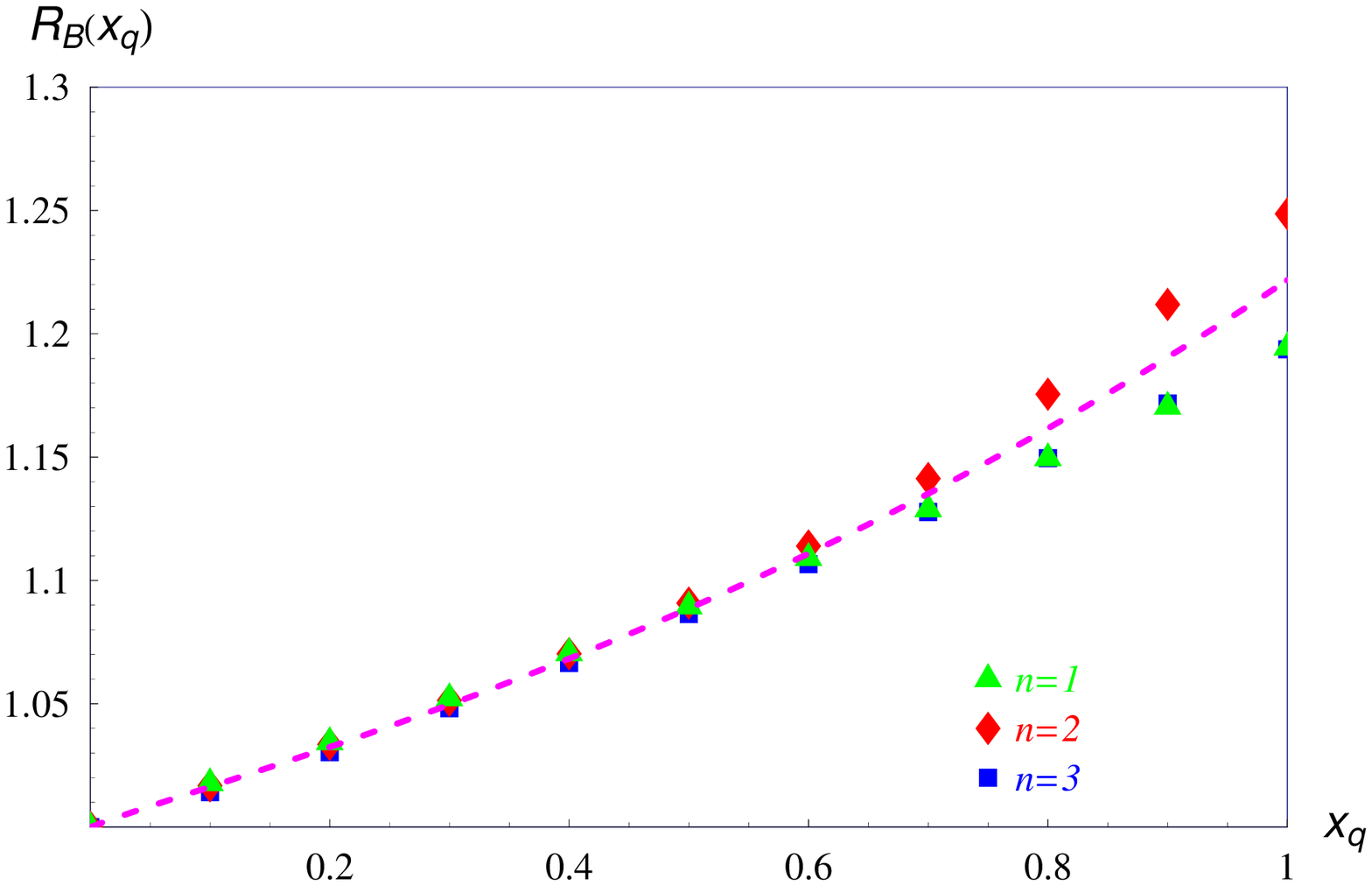} & \includegraphics[width=8cm]{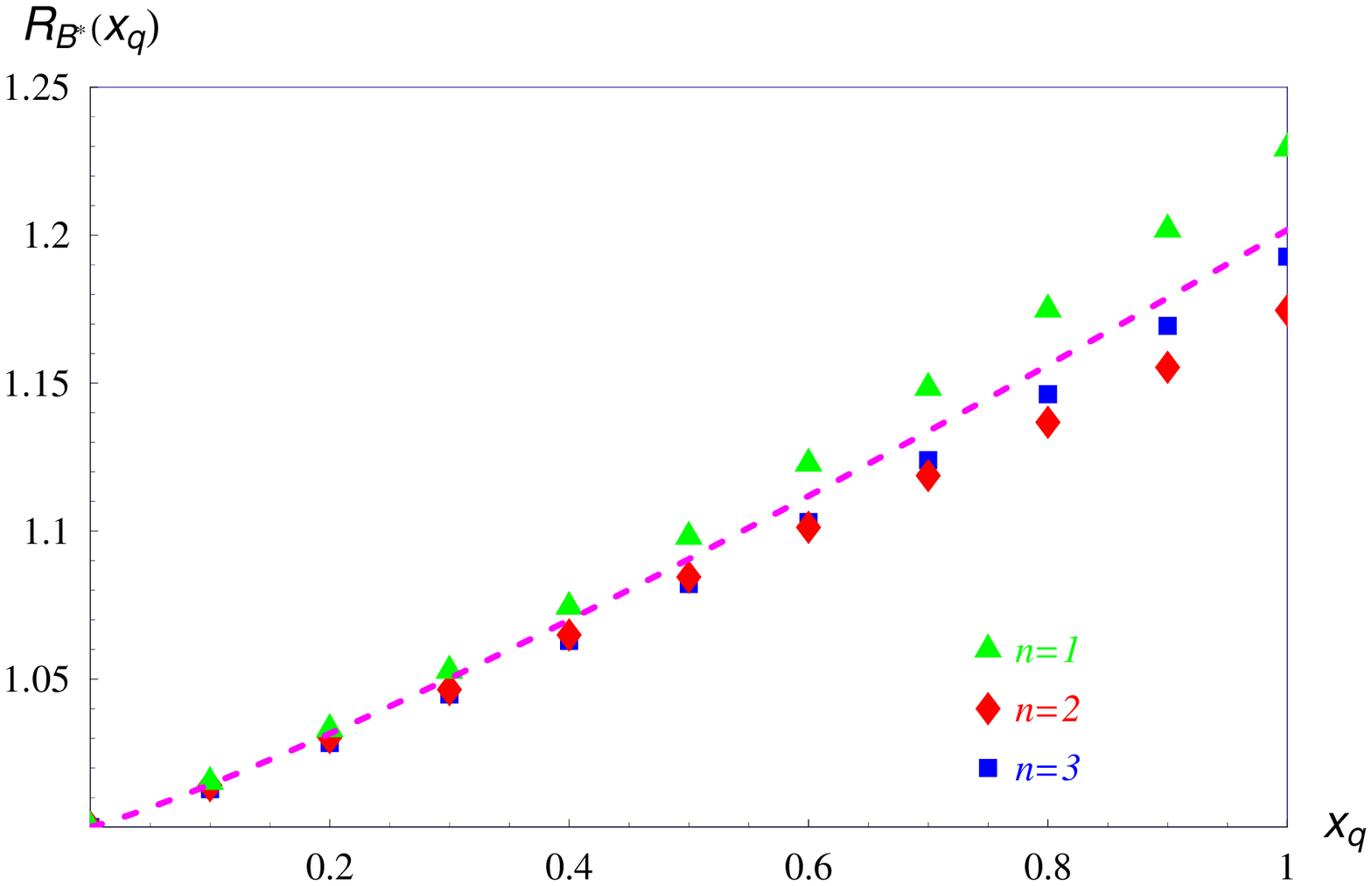} 
\end{tabular}
\caption{\label{Plot:1}
QCD sum-rule results for the ratio $R_{H_q}(x_q)$ (see Eq.~(\ref{RHq})) versus the variable $x_q$ (see Eq.~(\ref{xq})) for $H_q =\{D, D^*, B, B^*\}$, 
obtained using the central values of the OPE parameters given in Eq.~(\ref{Table:1}). The green triangles, red diamonds 
and blue squares correspond, 
respectively, to the use of the  linear $n=1$, quadratic $n=2$ and cubic $n=3$ Ans\"atze for the effective threshold $s_{\rm eff}^{H_q}$ 
(see Eq.~(\ref{seff})). The dashed lines correspond to the centers of the bands encompassed by the $n=1$, $n=2$, and $n=3$ results and 
provide our sum-rule estimates. The systematic errors are given by the half-widths of the bands.} 
\end{figure}

In order to reduce the uncertainty generated by the errors of the OPE parameters given in Eq.~(\ref{Table:1}), we consider the following 
ratio of decay constants
\begin{eqnarray}
\label{RHq}
R_{H_q}(x_q) \equiv f_{H_q}(x_q) / f_{H_{ud}}(0).
\end{eqnarray}
The function $R_{H_q}(x_q)$ is analytic near $x_q=0$ and has a Taylor expansion in some vicinity of this point: 
ChPT suggests the rightmost logarithmic singularity at $m_q=0$, i.e. at $x_q= - \frac{m_{ud}}{m_s - m_{ud}}$ (see Appendix).  
In order to reach the $H_{d/u}$ mesons, we need to set $x_q= (+/-)\frac{m_d -– m_u}{2(m_s -– m_{ud})}$, which both lie within the convergence radius 
of the Taylor expansion. Therefore, the slope $R_{H_q}^\prime(0)$ gives the leading IB effect on the heavy-meson decay constants 
\begin{eqnarray}
\label{deltaf_def}
\frac{f_{H_d} - f_{H_u}}{f_{H_{ud}}} =  R_{H_q}^\prime(0) \frac{m_d - m_u}{m_s - m_{ud}} ~ .
\end{eqnarray}
Figure \ref{Plot:1} shows our sum-rule results of the ratio $R_{H_q}(x_q)$ for $0 < x_q < 1$ in the case of $D$, $D^*$, $B$ and $B^*$ mesons.
Notice that our sum-rule results imply $f_{H_d} > f_{H_u}$ for $D$-, $D^*$-, $B$- and $B^*$-mesons.

The obtained numerical results in the range $0 < x_q < 1$ may be excellently described by a polynomial function of $x_q$; in practice, 
the quadratic fit is sufficient and adding higher powers of $x_q$ does not change the result, so we may write 
\begin{eqnarray}
\label{Rquad}
R_{H_q}(x_q) = 1 + R_{H_q}^{(1)} x_q + R_{H_q}^{(2)} x_q^2 ~ , 
\end{eqnarray}
where $R_{H_q}^{(1)}$ and $R_{H_q}^{(2)}$ are the parameters to be determined by the fit. On the basis of these results one obtains 
the IB effect in terms of the $R_{H_q}^{(1)}$. 

On the other hand, the Heavy Meson ChPT (HMChPT) predicts that the $m_q$-dependence of both pseudoscalar and vector heavy-meson decay 
constants contain chiral logs \cite{HMCHPT} which emerge as the nonperturbative effect of the soft region where light pseudoscalars 
play the crucial role. The nonperturbative physics contributes to our sum rule (\ref{srN}) 
in two ways: as power corrections and through the effective threshold. Respectively, the chiral logs may appear not only in the quark condensates 
\cite{qq-condensate} but also implicitly through the effective threshold.\footnote{For an interesting discussion of the chiral behaviour of QCD 
sum rules in the case of light hadrons, we refer to \cite{cohen}, where an alternative formulation of the sum rule with an explicit inclusion of 
the multipion states has been discussed. In principle, a treatment of heavy mesons along the same lines may be considered but this goes far beyond 
the scope of our interests here.} So one may doubt whether or not these chiral logs may alter the IB obtained on the basis of the polynomial fit 
(\ref{Rquad}). To check this, we use an extended Ansatz for $R_{H_q}(x_q)$ in the full interval $0 < x_q < 1$ which explicitly 
includes also the known chiral logs: 
\begin{eqnarray}
\label{Rchiral}
R_{H_q}(x_q) =1 + R_\chi(x_q) + \hat{R}_{H_q}^{(1)} x_q + \hat{R}_{H_q}^{(2)} x_q^2 ~ . 
\end{eqnarray}
Here $R_\chi(x_q)$ is the known function, explicitly given by Eq.~(\ref{Rchi}) in the Appendix, the 
same for all the mesons and $\hat{R}_{H_q}^{(1, 2)}$ are the parameters to be determined by the fit. 

The results obtained for the slope $R_{H_q}^\prime(0) \equiv [dR_{H_q}(x_q) / dx_q ]_{x_q = 0}$ adopting the fitting functions 
(\ref{Rquad}) and (\ref{Rchiral}) are presented in Table \ref{Table:R}. 
We stress that the reported errors come mainly from the systematic uncertainties related to the limited accuracy of the method 
of QCD sum rules, because the ratio $R(x_q)$ is almost independent of the OPE uncertainties, given in Eq.~(\ref{Table:1}), which 
cancel each other to a large extent in this ratio.

The last column in Table \ref{Table:R} represents our final estimate obtained as an average of the outcomes of the two fitting procedures; 
this averaging is performed according to the prescription given by Eq.~(28) of Ref.~\cite{ETMC_Nf211}. 
Taking into account the results of both fits reflects the uncertainty induced by the fitting procedures 
and leads to a more conservative estimate of the total uncertainties in the IB effects.

\begin{table}[!tb]
\caption{Results for the slope $R_{H_q}^\prime(0) = [dR_{H_q}(x_q) / dx_q ]_{x_q = 0}$ obtained by adopting 
the polynomial (\ref{Rquad}) and the HMChPT (\ref{Rchiral}) fitting functions in the full interval $0 < x_q < 1$. 
The last column represents the average of the two results (see text).}
\label{Table:R}
\vspace{2ex}
\begin{tabular}{|c|c|c|c|}
\hline
Meson   & \multicolumn{3}{c|}{$R_{H_q}^\prime(0)$}\\
\hline
              & polynomial fit &   HMChPT fit         &  average            \\
\hline \hline
$D$    &     $\qquad 0.144 \pm 0.009 \qquad$ &  $\qquad 0.171 \pm 0.009\qquad$   &   $\qquad 0.157 \pm 0.009\qquad$ \\ \hline
$D^*$  &     $0.218 \pm 0.021$ &  $0.248 \pm 0.022$   &   $0.233 \pm 0.026$ \\ \hline \hline           
$B$    &     $0.146 \pm 0.008$ &  $0.174 \pm 0.008$   &   $0.160 \pm 0.016$ \\ \hline
$B^*$  &     $0.139 \pm 0.009$ &  $0.162 \pm 0.010$   &   $0.150 \pm 0.015$ \\ \hline
\end{tabular}
\end{table}

Adopting $m_d - m_u ~ (2\;{\rm GeV}) =  (2.67 \pm 0.22)$ MeV from Ref.~\cite{FLAG}, one obtains
\begin{eqnarray}
\label{deltaf_res}
(f_{D^+} - f_{D^0}) / f_{D} & = & 0.0047 (6) ~ , \nonumber \\
(f_{D^{*+}} - f_{D^{*0}}) / f_{D^*} & = & 0.0068 (9) ~ , \nonumber \\
(f_{B^0} - f_{B^+}) / f_{B} & = & 0.0047 (6) ~ , \nonumber \\
(f_{B^{*0}} - f_{B^{*+}}) / f_{B^*} & = & 0.0045 (5) ~ ,  
\end{eqnarray}
which exhibit a nice accuracy of the order of $\simeq 10 - 15\%$.

The IB differences $f_{H_d} - f_{H_u}$ have clearly a bit worse accuracy, since they are influenced by the uncertainties of 
the OPE parameters and by the residual unphysical scale-dependence of the extracted decay constants related to the truncation of the OPE series.
Adopting the sum-rule values $f_D = 206.2 \pm 8.9$ MeV \cite{lms_fD+fDstar}, $f_{D^*} = 252.2 \pm 22.7$ MeV \cite{lms_fD+fDstar}, 
$f_B = 192.0 \pm 14.6$ MeV \cite{ourmb} and $f_{B^*} = 181.8 \pm 13.7$ MeV \cite{lms_fB_ratio}, we get
\begin{eqnarray}
\label{deltaf_D}
f_{D^+} - f_{D^0} & = & 0.97 \pm 0.13 ~ {\rm MeV} ~ , \\
\label{deltaf_Dstar}
f_{D^{*+}} - f_{D^{*0}} & = & 1.73 \pm 0.27 ~ {\rm MeV} ~ , \\ 
\label{deltaf_B}
f_{B^0} - f_{B^+} & = & 0.90 \pm 0.13 ~ {\rm MeV} ~ , \\
\label{deltaf_Bstar}
f_{B^{*0}} - f_{B^{*+}} & = & 0.81 \pm 0.11 ~ {\rm MeV} ~ ,
\end{eqnarray}
which show that the IB effect found in our analysis of the decay constants of pseudoscalar and vector heavy mesons has a typical 
size of $\approx 1$ MeV with an accuracy of $\approx 15 \%$.
In the case of $D$- and $B$-mesons the IB effect has been calculated recently using lattice QCD, 
namely $f_{D^+} - f_{D^0} = 0.94_{-0.12}^{+0.50}$ MeV \cite{lattice_IB_fD} and $f_{B^0} - f_{B^+} = 3.8 \pm 1.0$ MeV \cite{lattice_IB_fB,PDG16_fPS}.
Thus in the case of the $D$-meson the lattice QCD \cite{lattice_IB_fD} and our sum-rule (\ref{deltaf_D}) 
results are nicely consistent with each other, whereas in the case of the $B$-meson the lattice QCD result \cite{lattice_IB_fB,PDG16_fPS} 
is almost $4$ times larger than our sum-rule one (\ref{deltaf_B}) with a tension of $\approx 3$ standard deviations.

We close this section by noting that the IB effect in $K$-meson, $f_{K^0} - f_{K^+}$, has been recently calculated on the lattice in 
Ref.~\cite{lattice_IB_fK}, obtaining $f_{K^0} - f_{K^+} = 1.24 \pm 0.59$ MeV.
Thus the size of the IB effect in $K$-meson appears to be similar to the one found in our analysis for the $D$-, $D^*$-, $B$- and $B^*$-mesons.

\section{Summary and conclusions}
We have studied the IB effects in the decay constants of pseudoscalar and vector heavy-light mesons using the method of QCD sum rules. 
For vector mesons our results provide the first estimates available in the literature. 

We made use of the axial-vector and vector correlation functions in which we took into account the full light-quark mass dependence 
in the LO perturbative spectral density, $O(m_q^0 \alpha_s)$, $O(m_q \alpha_s)$ and $O(m_q^0 \alpha_s^2)$ radiative corrections.
Knowing the explicit dependence of the OPE on the light-quark mass and obtaining the decay constants for various values of the 
light-quark masses in the range from the average $u/d$ quark mass to the strange quark mass opens the possibility to access the 
IB effects at first order in the quark mass difference $m_d - m_u$. 

We report the following results: 
\begin{itemize}
\item The ratios $[f_{H_d} - f_{H_u}] / f_{H_{ud}}$ are predicted with rather good accuracy as they are almost free from the OPE 
uncertainties which cancel each other in these ratios. 
Adopting $m_d - m_u ~ (2\;{\rm GeV}) =  (2.67 \pm 0.22)$ MeV \cite{FLAG} we obtain 
\begin{eqnarray}
(f_{D^+} - f_{D^0}) / f_{D} & = & 0.0047 (6) ~ , \nonumber \\
(f_{D^{*+}} - f_{D^{*0}}) / f_{D^*} & = & 0.0068 (9) ~ , \nonumber \\ 
(f_{B^0} - f_{B^+}) / f_{B} & = & 0.0047 (6) ~ , \nonumber \\
(f_{B^{*0}} - f_{B^{*+}}) / f_{B^*} & = & 0.0045 (5) ~ , 
\end{eqnarray}
where the quoted uncertainties are dominated by the systematic uncertainties of the sum-rule extraction. 
\item The IB differences of the decay constants have a bit worse accuracy, as they are influenced by the uncertainties of the 
OPE parameters (see Eq.~(\ref{Table:1})) and by the residual unphysical scale-dependence of the extracted decay constants, 
related to the truncation of the OPE series. 
For the IB differences $f_{H_d} - f_{H_u}$ for $H = \{D, D^*, B, B^*\}$ we get 
\begin{eqnarray}
f_{D^+} - f_{D^0} & = & 0.97 \pm 0.13 ~ {\rm MeV} ~ , \\
f_{D^{*+}} - f_{D^{*0}} & = & 1.73 \pm 0.27 ~ {\rm MeV} ~ , \\
f_{B^0} - f_{B^+} & = & 0.90 \pm 0.13 ~ {\rm MeV} ~ , \\
f_{B^{*0}} - f_{B^{*+}} & = & 0.81 \pm 0.11 ~ {\rm MeV} ~ , 
\end{eqnarray}
showing that the size of the leading IB effect is $\approx 1$ MeV and that $f_{H_d} > f_{H_u}$.
\end{itemize}

We stress that in the case of the $D$-meson the lattice QCD \cite{lattice_IB_fD} and our sum-rule results are nicely consistent 
with each other, whereas in the case of the $B$-meson the lattice QCD result \cite{lattice_IB_fB,PDG16_fPS} is almost $4$ times 
larger than our sum-rule one with a tension of $\approx 3$ standard deviations.

\appendix
\section{HMChPT prediction for the ratio $R_{H_q}$}
The partially quenched ChPT for heavy-light mesons coupled to pions and kaons was formulated in Ref.~\cite{HMCHPT} to calculate 
the one-loop chiral logs occurring in the heavy-meson decay constant in the heavy-quark limit $m_Q \to \infty$.
In terms of the quantity $\Phi_{H_q} \equiv f_{H_q} \sqrt{M_{H_q}}$ one has
\begin{eqnarray}
\label{PhiHq}
\Phi_{H_q} & = & \Phi \left\{ 1 + C_1(\nu) (2 \chi_{ud} + \chi_s) + C_2(\nu) \chi_q - \frac{1 + 3 \hat{g}^2}{(4 \pi f_0)^2} \left[ 
                            \frac{\chi_{ud} + \chi_q}{2} \mbox{log}\left(\frac{\chi_{ud} + \chi_q}{2 \nu^2}\right) \right. \right. \nonumber \\
                  & + & \left. \left.  \frac{\chi_s + \chi_q}{4} \mbox{log}\left(\frac{\chi_s + \chi_q}{2 \nu^2}\right) - \frac{1}{2} 
                            \frac{2(\chi_s - \chi_q)^2 + (\chi_{ud} - \chi_q)^2}{(2\chi_s + \chi_{ud} - 3 \chi_q)^2} \chi_q 
                            \mbox{log}\left(\frac{\chi_q}{\nu^2}\right) \right. \right. \nonumber \\
                  & + & \left. \left. \frac{1}{3} \frac{(\chi_s - \chi_{ud})^2}{(2\chi_s + \chi_{ud} - 3 \chi_q)^2} \frac{\chi_{ud} + 2 \chi_s}{3} 
                            \mbox{log}\left(\frac{\chi_{ud} + 2 \chi_s}{3 \nu^2}\right) \right. \right. \nonumber \\
                  & - & \left. \left. \frac{\chi_q - \chi_{ud}}{2} \frac{\chi_s - \chi_q}{2\chi_s + \chi_{ud} - 3 \chi_q} 
                           \left( 1+ \mbox{log}\left(\frac{\chi_q}{\nu^2}\right) \right) \right] \right\} ~ ,
\end{eqnarray}
where $\chi_i \equiv 2 B_0 m_i$ ($i = q, ud, s$), $m_{ud}$ and $m_s$ are the average $u/d$ and strange sea-quark masses, 
respectively, $m_q$ is the valence light-quark mass, and $C_1(\nu)$ and $C_2(\nu)$ are low energy constants (LECs) depending 
on the ChPT renormalization scale $\nu$. 
Such a dependence is cancelled by the corresponding one in the chiral logs, so that $\Phi_{H_q}$ is independent on the scale $\nu$.
In Eq.~(\ref{PhiHq}) the parameter $\hat{g}$ is related to the strong coupling constant $g_{V P \pi}$ by 
$g_{V P \pi} = 2 \hat{g} \sqrt{M_P M_V} / f_\pi$ 
and is chosen to be equal to the Heavy Quark Effective Theory result $\hat{g} = 0.44 (8)$ \cite{HQET}.
Finally the parameters $B_0$ and $f_0$ are the LECs of the SU(3) ChPT at LO.

For $m_q = m_{ud}$ one gets
\begin{eqnarray}
\label{PhiHud}
\Phi_{H_{ud}} & = & \Phi \left\{ 1 + C_1(\nu) (2 \chi_{ud} + \chi_s) + C_2(\nu) \chi_{ud} - \frac{1 + 3 \hat{g}^2}{(4 \pi f_0)^2} \left[ 
                                \frac{3}{4} \chi_{ud} \mbox{log}\left(\frac{\chi_{ud}}{\nu^2}\right) \right. \right. \nonumber \\
                      & + & \left. \left.  \frac{\chi_s + \chi_{ud}}{4} \mbox{log}\left(\frac{\chi_s + \chi_{ud}}{2 \nu^2}\right) + 
                                \frac{\chi_{ud} + 2 \chi_s}{36} \mbox{log}\left(\frac{\chi_{ud} + 2 \chi_s}{3 \nu^2}\right) \right] \right\} ~ .
\end{eqnarray}
Therefore, for the chiral logs in the ratio $R_{H_q}$, given by Eq.~(\ref{Rchiral}), one obtains (choosing $\nu^2 = 2 B_0 m_s$)
\begin{eqnarray}
\label{Rchi}
R_\chi(x_q) & = & - 2 B_0 \frac{1 + 3 \hat{g}^2}{(4 \pi f_0)^2} \left\{  \frac{m_{ud} + m_q}{2} 
                                                     \mbox{log}\left(\frac{m_{ud} + m_q}{2 m_s}\right) + \frac{m_s + m_q}{4} 
                                                     \mbox{log}\left(\frac{m_s + m_q}{2 m_s}\right) \right. \nonumber \\
                                           & - & \left. \frac{1}{2} \frac{2(m_s - m_q)^2 + (m_{ud} - m_q)^2}{(2m_s + m_{ud} - 3 m_q)^2} m_q 
                                                    \mbox{log}\left(\frac{m_q}{m_s}\right) - \frac{3}{4} m_{ud} 
                                                    \mbox{log}\left(\frac{m_{ud}}{m_s}\right) \right. \nonumber \\
                                           & + & \left.  \frac{1}{3} \left[ \frac{(m_s - m_{ud})^2}{(2 m_s + m_{ud} - 3 m_q)^2}- \frac{1}{4} \right] 
                                                     \frac{m_{ud} + 2 m_s}{3} \mbox{log}\left(\frac{m_{ud} + 2 m_s}{3 m_s}\right) \right. \nonumber \\
                                           & - & \left. \frac{m_q - m_{ud}}{2} \frac{m_s - m_q}{2 m_s + m_{ud} - 3 m_q} 
                                                    \left( 1+ \mbox{log}\left(\frac{m_q}{m_s}\right) \right)- \frac{m_{ud} + m_s}{4}  
                                                    \mbox{log}\left(\frac{m_{ud} + m_s}{2 m_s}\right) \right\} ~ ,
\end{eqnarray}
where $m_q=m_{ud}x_q+m_s(1-x_q)$. Notice that $R_\chi(0)=0$. 

\section*{\bf Acknowledgments}
D.M.~was supported by the Austrian Science Fund (FWF) under project P29028. S.S.~warmly thanks S.R.~Sharpe for providing 
the extension of the calculation of the chiral logs of Ref.~\cite{HMCHPT} to the case of $N_f = 2 + 1$ dynamical quarks.

\end{document}